\RequirePackage{fix-cm}
\pdfoutput=1
\documentclass[epjc3]{svjour3}
\smartqed  
\RequirePackage{graphicx}

\usepackage[margin=2.0cm]{geometry}

\usepackage{amssymb}
\usepackage[numbers]{natbib}
\journalname{Eur. Phys. J. C}

\usepackage{latexsym}
\usepackage{revsymb}
\usepackage{multirow}
\usepackage{color}
\usepackage[usenames,dvipsnames,svgnames,table]{xcolor}

\usepackage{graphicx}
\usepackage{epsfig}  
\usepackage{epsf}    
\usepackage{dcolumn}
\usepackage{bm}
\usepackage{dcolumn}
\usepackage{textcomp}
\usepackage{float}
\usepackage{subfig}

\usepackage{lineno}
\modulolinenumbers[5]

\usepackage[bookmarks=false]{hyperref}
\usepackage[all]{hypcap}
\usepackage{pdftricks}
\usepackage{amsmath}
\usepackage{bm}
\usepackage{amssymb}
\usepackage{amsfonts}

\begin{document}




\title{Potential impact of sub-structure on the resolution of neutrino mass hierarchy at medium-baseline reactor neutrino oscillation experiments}
\author{Zhaokan Cheng\thanksref{e1,addr1}
        \and
        Neill Raper\thanksref{e2,addr2}
        \and
        Wei Wang\thanksref{e3,addr2}
        \and
        Chan Fai Wong\thanksref{e4,addr2}
        \and
        Jingbo Zhang\thanksref{e5,addr1}
}

\thankstext{e1}{14b911018@hit.edu.cn}
\thankstext{e2}{palebluedot89@gmail.com}
\thankstext{e3}{wangw223@sysu.edu.cn}
\thankstext{e4}{wongchf@sysu.edu.cn}
\thankstext{e5}{jinux@hit.edu.com}


\institute{School of Physics, Harbin Institute of Technology, Harbin, China \label{addr1}
           \and
           School of Physics, Sun Yat-sen University, GuangZhou, China \label{addr2}
}

\date{}

\maketitle

\begin{abstract}
In the past decade, the precise measurement of the lastly known
neutrino mixing angle $\theta_{13}$ has enabled the potential resolution of
neutrino mass hierarchy (MH) at medium-baseline reactor neutrino oscillation (MBRO) experiments.
On the other hand, recent calculations of the reactor
neutrino flux predict percent-level sub-structures in the $\bar\nu_e$
spectrum due to Coulomb effects in beta decay. Such
fine structure in the reactor spectrum could be an important issue for
the resolution of neutrino MH for the MBRO approach since they could affect the sub-dominant oscillation
pattern used to discriminate different hierarchies.
Inconveniently, the energy resolutions of current reactor experiments are not sufficient to measure such fine
structure, and therefore the size and location in energy of these
predicted discontinuities has not been confirmed experimentally.
There has been speculation that a near detector is required with sufficient
energy resolution to resolve the fine structure such that it can be
accounted for in any analysis which attempts to discriminate the
MH. This article studies the impact of fine structure on the
resolution of MH, based on the predicted reactor neutrino
spectra, using the measured spectrum from Daya Bay as a reference. We
also investigate whether a near detector could improve
the sensitivity of neutrino MH resolution using various
assumptions of near detector energy resolution.


\end{abstract}
\maketitle


\pagenumbering{arabic}

\section{Introduction}\label{sec1}
The most challenging neutrino mixing angle $\theta_{13}$ has been
measured precisely and was found to be larger than previously expected
by the current generation of short-baseline reactor neutrino
and long-baseline accelerator neutrino experiments
\cite{An:2012eh, Ahn:2012nd, Abe:2011fz, Abe:2013hdq, Adey:2018zwh}. The
unexpectedly large value of $\theta_{13}$ allows for the measurement
of a leptonic CP-violating phase and the resolution of neutrino
mass hierarchy (MH). Particularly, it opens a gateway to determine the
MH from (approximately) vacuum oscillation in medium-baseline reactor
neutrino oscillation (MBRO) experiments\cite{Petcov, Zhan:2008id,
  Zhan:2009rs, Qian:2012xh, Kettell:2013eos, Takaesu:2013wca,
  Li:2013zyd, Kim:2014rfa, An:2015jdp, Chan:2015mca}.
These experiments are designed to resolve the neutrino MH via precise
spectral measurement of reactor $\bar\nu_e$ oscillations. A large
liquid-scintillator detector ($\sim$ 10 - 20 ktons) with excellent
energy resolution (3\%/$\sqrt{E/MeV}$), located $\sim$50 km from the
reactor core(s) is expected to be able to observe the sub-dominant
oscillation pattern and thus discriminate the MH by measuring the
resulting spectral distortions \cite{Kim:2014rfa, An:2015jdp}.
However it is possible that the fine structure predicted to exist in
the reactor neutrino spectrum might constructively or destructively
interfere with the spectral distortions used to determine the MH.

In parallel, reactor neutrino experiments
have also measured reactor antineutrino flux and spectrum with
unprecedented statistics at distances from dozens of meters to $\sim$2km from reactor sources.
Together with the results from previous experiments, we had found a deficit in reactor neutrinos relative
to predictions, known as the reactor antineutrino anomaly (RAA) \cite{Anomaly}.
After counting the deficit, current experiments found an excess of
events with respect to predictions in the region of 4 to 6 MeV prompt
energy, which came to be known as the ``bump'' or
``shoulder''\cite{An:2013zwz, Seo:2014xei, Ko:2016owz}.
Recently, it was also found that predictions of the (unoscillated)
reactor antineutrino spectrum \cite{Huber:2011wv, Mueller:2011nm} is inconsistent with
the latest experimental measurements in the ratio between $^{235}$U
and $^{239}$Pu yields \cite{Hayes:2017res, An:2017osx, Adey:2019ywk}.
In addition to the larger scale shape discrepancy, attempts
of predicting reactor antineutrino flux and spectrum using \textit{ab
  initio} approaches predict
percent-level sub-structures in the $\bar\nu_e$ spectrum due to
Coulomb effects in beta decay\cite{Dwyer:2014eka, Sonzogni:2017voo}.
Since the sensitivity to neutrino MH in the MBRO experiments
strongly depends on measurements of the antineutrino survival spectrum
and its uncertainties, this undetermined
reactor antineutrino spectrum at the same scale could be an important issue for the
resolution of neutrino MH\cite{Capozzi:2015bpa}.

In this context, we present numerical simulations to investigate the potential impact of fine structure in the reactor neutrino flux on the resolution of neutrino MH.
We start with the simplest arrangement for a MBRO experiment, namely, one powerful source and one single large detector with a baseline of 52.5 km.
We also investigate whether a near detector can provide significant
improvement on the MH sensitivity, and the effect on this sensitivity
from different values of near detector energy resolution.

This paper is organized as follows. In section \ref{sec2}, we review the discrepancies between reactor flux measurements and conventional predictions. Then, in section \ref{sec3}, we present our estimation of the scale of fine structure and our simulation of MH resolution sensitivity, with additional shape uncertainties due to fine structure taken into account.
In section \ref{sec4}, we present our study concerning the proposed near detector. Finally, a summary of our results and perspectives are concluded in section \ref{sec5}.

\section{The Undetermined Reactor Spectrum and Discrepancies Between Experiments and Conventional Predictions}\label{sec2}
In previous experiments, the conventional method of predicting reactor neutrino flux uses measurements of electron spectra from the beta decays of fission daughters.
These spectra are fit with fake beta branches from high energy bins to low, subtracting each ``virtual branch'' spectrum before fitting the next energy bin \cite{Huber:2011wv}. Those virtual branches are then converted to $\bar\nu_e$ spectra via the relation $E_{0} - E_{\bar\nu_e} = T_{\beta}$, where $E_{0}$ is the available energy for the beta decay.
Beta-conversion antineutrino spectra of $^{235}$U, $^{239}$Pu, $^{241}$Pu from Huber \cite{Huber:2011wv} and $^{238}$U spectra from Mueller \cite{Mueller:2011nm} have been the most widely-used in reactor neutrino experiments.
The conversion method was favored because corresponding uncertainties were well-defined and associated with the conversion procedure.
Such predictions of $\beta$ decay spectrum estimate that the uncertainties of $\bar\nu_e$ would be around a few percent. However, the shoulder in the reactor neutrino spectrum \cite{Hayes:2015yka} and reactor neutrino anomaly \cite{Anomaly} along with the recent discovery that Huber-Mueller estimates the $^{235}$U spectrum particularly poorly indicate that these uncertainties are not correct and conversion spectra are not as reliable as once thought.

The only other method of predicting reactor neutrino spectra is the summation method described in \cite{PhysRevC.91.011301,Dwyer:2014eka,Hayes:2015yka,Sonzogni:2017voo,Hayes:2017res}. To generate a summation prediction, one first calculates beta decay spectrum for every contributing isotope. Following that normalize each total beta spectrum to the cumulative yield of its corresponding isotope. The cumulative yield ($Y_c$) is the probability that the isotope appears as a result of either a fission, or the decay of the other fission products, and therefore represents the fraction of the reactor neutrino spectrum which decays from that isotope will contribute.
In principle, the result should be the true reactor neutrino spectrum though still not exact due to some approximations used in correcting for various effects. However, the dominant uncertainty and bias comes from the underlying data (Q values, transition probabilities, energy levels, and cumulative fission yields). Besides the uncertainties, the bias in the underlying measurements is a larger concern for the method. The most well understood of these is the pandemonium effect \cite{Hardy:1977suw}, which overestimates feeding to lower energy levels. Recent experiments conducted by the IGISOL collaboration have addressed this bias for a few isotopes, which contribute strongly to the reactor neutrino spectrum \cite{Zakari-Issoufou:2015vvp,Rice:2017kfj,Valencia:2016rlr}, and these results are included in our analysis.
However, additional biases likely remain as well as results which suffer from the pandemonium effect, but have not yet had new experiments measure the structure data to greater precision.

The majority of fission isotopes in Pressurized Water Reactors (PWRs) are $^{235}$U and $^{239}$Pu. As discussed above, the conversion method predicts the $^{239}$Pu rate more accurately than the summation method, but predicts the $^{235}$U spectrum poorly. Whereas the summation method overestimates the flux from both, but predicts the flux ratio fairly well \cite{Hayes:2017res}. At recent, further updated summation calculations have found better agreement with overall flux \cite{Estienne:2019ujo}.
However, regardless of the quality of overall flux prediction, we are forced to use the summation method in this analysis because we aim to estimate the fine structure in the spectrum. Each beta decay has a sharp discontinuity at its endpoint. When all the individual beta decay spectra are summed fine structure emerges. Because the conversation method uses fake beta branches, fine structure can only be predicted by the summation method, therefore the summation method is used exclusively in this paper.

The origin of the discontinuities which give rise to fine structure is as follows. The beta spectrum from the beta decay of a free neutron will go to zero at zero kinetic energy. However, when an isotope decays into another, the electron is generated quite close to the positively charged nucleus. Therefore, the effect of this charge on the negatively charged electron needs to be accounted for in the available final states. This results in a relatively soft electron energy spectrum, which importantly does not go to zero at zero kinetic energy.
If we recall the relationship $E_{0} = T_{\beta} + E_{\bar\nu_e}$, which tells us that if many electrons are created with zero kinetic energy, we should also see many neutrinos created with an energy equal to $E_{0}$, the maximum available energy. This implies a sharp discontinuity at the endpoint of each neutrino spectrum corresponding to each beta decay branch. When the neutrino spectra of these beta branches are summed together, the discontinuities at the end of each generate fine structure in the total reactor neutrino spectrum, which is found to be at the few percent level \cite{priv:Langford}, as shown in reference \cite{Dwyer:2014eka}. The size of the fine structure is dominated by the relative Fission Yields between isotopes. Large buildups of discontinuities also occur which can combine together to form what appears to be a single large discontinuity. More details and discussions of fine structure in reactor spectra can be found in references \cite{priv:Langford, Sonzogni:2017voo, Littlejohn:2018hqm}.

Recent short baseline reactor neutrino experiments such as Daya Bay, cannot measure fine structure as their detectors have only $\approx$ 8\% energy resolution. \footnote{However a bin to bin measurement has been attempted in order to attempt to measure a large buildup of discontinuities \cite{Sonzogni:2017voo}.}
However, the future MBRO experiments such as JUNO\cite{An:2015jdp} and RENO-50 \cite{Kim:2014rfa}, are expected to have finer detector energy resolution (around 3\%) and thus are expected to observe more fine structure in the spectrum.
However, it has been speculated that undetermined fine structure could give rise to unexpected variation in the measured spectrum which happens to mimic one or another MH, and therefore either reinforce the correct MH, or wash out its signal, resulting in either an artificially low or artificially high sensitivity.
For this to be true the fine structure would need to be at approximately the scale of the sub-dominant oscillation pattern, and would have to fall in the right places and at the right magnitudes to mimic the MH signal.
In the following, we treat the scale of the fine structure as an additional shape uncertainty because we expect to see fine structure but do not want our analysis to assume that the fine structures exact shape is predicted perfectly.
The next section will discuss the impact of fine structure on MH measurements.

\section{The Potential Impact of Fine Structure on the Resolution of Neutrino MH}\label{sec3}
\subsection{The Conventional Simulations of the MBRO Experiments}
\begin{figure}[!htbp]
\centering
  \includegraphics[scale=0.6]{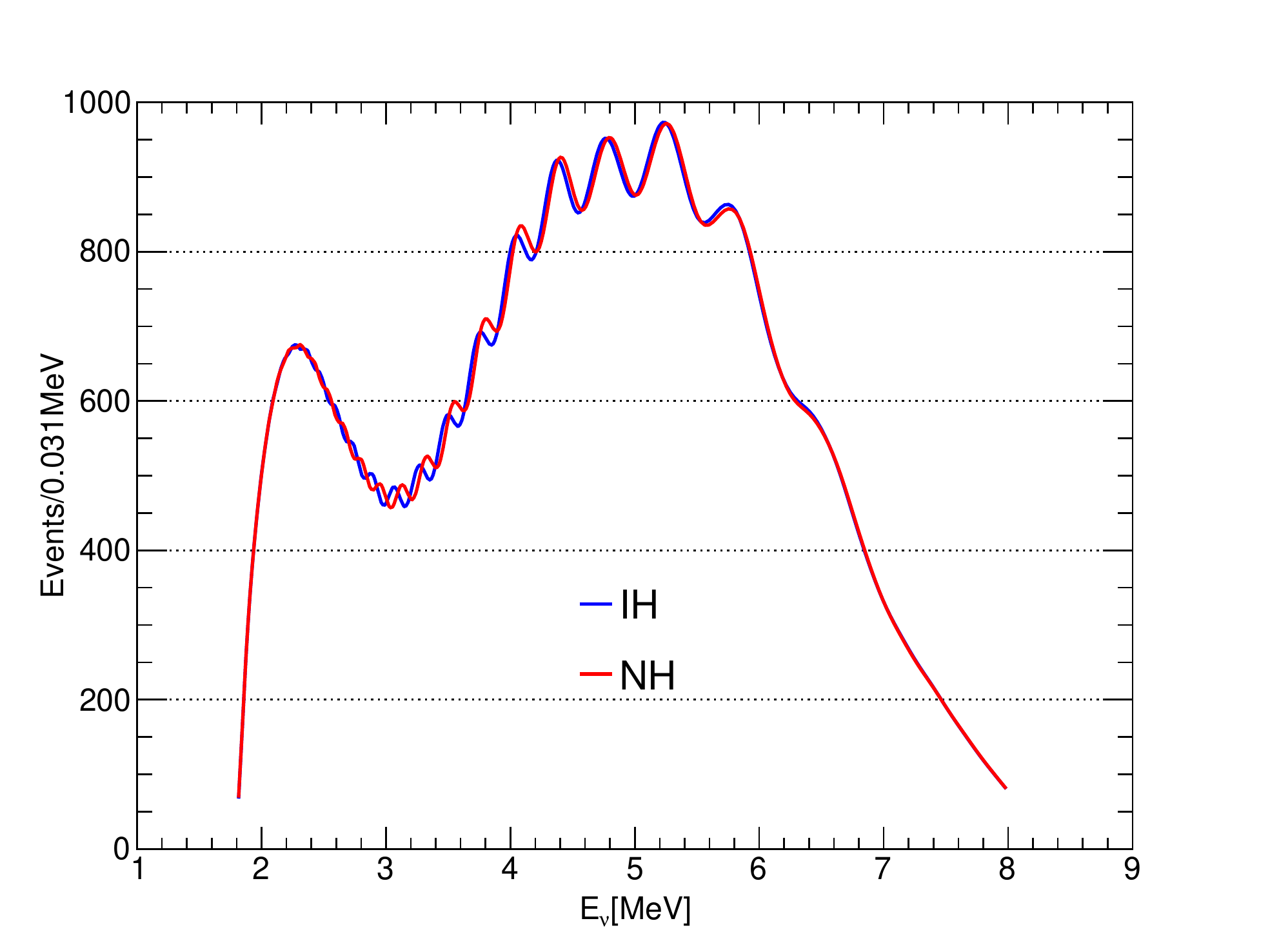}
 \caption{Expected reactor neutrino energy spectrum observed by a detector with 3\% energy resolution locating at 52.5 km. The red (blue) curve corresponds to the NH(IH) assumption. A medium-baseline reactor neutrino oscillation experiment(s) is expected to observe the subdominant $\theta_{13}$ oscillation and distinguish the tiny differences between the blue and red curves \cite{An:2015jdp, Kim:2014rfa}.}
 \label{NHIHspectrum}
\end{figure}

Future MBRO experiments are expected to identify the neutrino MH with $\Delta \chi^2$ $>$ 9. A large liquid-scintillator detector (20 ktons) is expected to be able to observe the sub-dominant oscillation pattern and thus extract the MH signal from the predicted spectral distortion, as illustrated in Fig. \ref{NHIHspectrum}.
To distinguish between normal hierarchy (NH) and inverted hierarchy (IH), we quantify the sensitivity of MH resolution by employing a least-squares method, calculating the difference between the (assumed) true event rate and the fitting event rates.

In our simulation, we assume the ideal detector is 20 ktons with 3\% energy resolution, located 52.5 km from the reactor core at medium-baseline reactor neutrino oscillation experiment(s). The electron antineutrino survival probability is given by \cite{Qian:2012xh, Li:2013zyd, An:2015jdp}
 \begin{align}\label{Eq_Pee}
    P_{\bar{e}\bar{e}} = & 1 - \mathrm{cos}^4(\theta_{13})\mathrm{sin}^2(2\theta_{12})\mathrm{sin}^2(\dfrac{\Delta m^2_{21} L}{4E}) - \mathrm{sin}^2(2\theta_{13})[\mathrm{cos}^2(\theta_{12})\mathrm{sin}^2(\dfrac{\Delta m^2_{31} L}{4E}) + \mathrm{sin}^2(\theta_{12})\mathrm{sin}^2(\dfrac{\Delta m^2_{32} L}{4E})] \notag \\
    = & 1 - \mathrm{cos}^4(\theta_{13})\mathrm{sin}^2(2\theta_{12})\mathrm{sin}^2(\dfrac{\Delta m^2_{21} L}{4E}) \notag \\
    & - \dfrac{1}{2}\mathrm{sin}^2(2\theta_{13}) [1 - \sqrt{1-\mathrm{sin}^2(2\theta_{12})\mathrm{sin}^2(\dfrac{\Delta m^2_{21} L}{4E})}\mathrm{cos}(\dfrac{2|\Delta m^2_{ee}L|}{4E}\pm \phi)].
\end{align}
where $\Delta m^2_{ee}$ is the effective mass-squared difference \cite{deGouvea:2005hk, Nunokawa:2005nx}. The values of $\Delta m^2_{ee}$ and $\phi$ in Eq. (\ref{Eq_Pee}) are given by
\begin{align}\label{Eq_Pee_explain}
 \Delta m^2_{ee} & = \mathrm{cos}^2 \theta_{12} \Delta m^2_{31} + \mathrm{sin}^2 \theta_{12} \Delta m^2_{32}, \\
 \mathrm{sin}\phi & = \dfrac{\mathrm{cos}^2 \theta_{12} \mathrm{sin}(2\mathrm{sin}^2 \theta_{12}\frac{\Delta m^2_{21} L}{4E}) - \mathrm{sin}^2 \theta_{12} \mathrm{sin}(2\mathrm{cos}^2 \theta_{12}\frac{\Delta m^2_{21} L}{4E})}{\sqrt{1-\mathrm{sin}^2(2\theta_{12})\mathrm{sin}^2(\frac{\Delta m^2_{21} L}{4E})}}, \\
 \mathrm{cos}\phi & = \dfrac{\mathrm{cos}^2 \theta_{12} \mathrm{cos}(2\mathrm{sin}^2 \theta_{12}\frac{\Delta m^2_{21} L}{4E}) + \mathrm{sin}^2 \theta_{12} \mathrm{cos}(2\mathrm{cos}^2 \theta_{12}\frac{\Delta m^2_{21} L}{4E})}{\sqrt{1-\mathrm{sin}^2(2\theta_{12})\mathrm{sin}^2(\frac{\Delta m^2_{21} L}{4E})}}.
\end{align}
In Eq. (\ref{Eq_Pee}), the positive sign corresponds to NH and negative sign corresponds to IH respectively. In our simulations, the values of the oscillation parameters are assumed to be 
$\Delta m^2_{21}$ = 7.59 $\times$ 10$^{-5}$ eV$^2$, $\Delta m^2_{ee}$ = 2.47 $\times$ 10$^{-3}$ eV$^2$, sin$^2 2\theta_{12}$ = 0.846, sin$^2 2\theta_{13}$ = 0.085 \cite{Wang:2016vua, Patrignani:2016xqp}.

We quantify the sensitivity of the MH measurement by employing the least-squares method, based on a $\chi^2$ function given by:
\begin{align}\label{chi2}
 \chi^2 & = \sum_i^{N_\mathrm{bins}} \dfrac{[T_i - F_i(1 + \eta_R + \eta_d + \eta_i)]^2}{T_i} + (\dfrac{\eta_R}{\sigma_R})^2 + (\dfrac{\eta_d}{\sigma_d})^2 + \sum_i^{N_\mathrm{bins}}(\dfrac{\eta_i}{\sigma_{s,i}})^2
\end{align}
where $T_i$ is measured neutrino event in the $i$th energy bin, and $F_i$ is the predicted number of neutrino events with oscillations taken into account (the fitting event rate). $\eta$ with different subscripts are nuisance parameters corresponding to reactor-related uncertainty ($\sigma_R$), detector-related uncertainty ($\sigma_d$) and shape uncertainty ($\sigma_s$). In references \cite{Yifang, An:2015jdp, Wang:2016vua}, $\sigma_R$ is assumed to be 2\% and $\sigma_d$ is assumed to be 1\%. Since MH sensitivity mainly depends on spectrum shape uncertainties, these uncertainties are minor and we follow the same assumptions in our analysis.
Shape uncertainties ($\sigma_{s,i}$) are modified by adding the scale of potential substructure in the spectrum as an additional shape uncertainty. Because these uncertainties are crucial to calculating the sensitivity of an MH measurement, we modify the assumption from other references for the shape uncertainty (which assume $\sigma_s$ to be constant at 1\%). We treat the shape uncertainty as energy dependent, namely, $\sigma_s$ = $\sigma_s (E_\nu)$.

Without loss of generality, in our simulations, the NH is assumed to be the true MH. The number of bins used $N_\mathrm{bins}$ is 200, equally spaced between 1.8 and 8 MeV. In this article, we focus on the potential impact of unknown structure in the reactor neutrino flux, and neglect the potential impact of detector nonlinearity \cite{Yifang, An:2015jdp}, actual reactor distribution \cite{Yifang, An:2015jdp}, and matter effects \cite{Li:2016txk}. The capability to resolve the MH is then given
by the difference between the minimum $\chi^2$ value for IH and NH:
\begin{equation}\label{delta_chi2}
 \Delta \chi^2 \equiv |\chi^2_\mathrm{min} (\text{IH}) - \chi^2_\mathrm{min} (\text{NH})|
\end{equation}

In the next two subsections, we will focus on treatment of the fine structure as an additional shape uncertainty and evaluate its effects on the sensitivity of MH resolution.

\subsection{Our Analysis Method}
\subsubsection{Estimation of Fine Structure}

To study the impact of spectrum fine structure on the resolution of neutrino MH, we estimate the scale of such structure (especially in the energy range between $E_\nu$ = 2 MeV to 6 MeV), and investigate how this additional shape uncertainty affects the $\Delta\chi^2$. We define the scale of the fine structure relative to a hypothetical measured spectrum smeared out according to a set energy resolution. In particular, this scale is determined by taking the ratio of the jagged spectrum to the smeared out spectrum.
Therefore, the scale of fine structure can only be determined with reference to a choice of energy resolution for the smeared spectrum. For the purposes of this analysis, we choose 8\%, which is the energy resolution of the Daya Bay detectors. We take the measured spectrum from Daya Bay, which is currently the most precise reactor neutrino measurement, as the nominal (unoscillated) spectrum in our simulation. We start with the measured spectrum from Daya Bay (26 bins) \cite{An:2016srz} and obtain a smooth spectrum, then estimate the sensitivity of MH resolution by calculating Eq. (\ref{chi2}) and Eq. (\ref{delta_chi2}) with this smooth Daya Bay spectrum used to estimate both $T_i$ and $F_i$.
Of course, fine structure is absent due to the finite (8\%) detector energy resolution, so we need to estimate the scale of this missing fine structure and add it to Eq. (\ref{chi2}) as an additional shape uncertainty.

In this case, fine structure is a potentially fuzzy term. What we really mean when we refer to the scale of the fine structure is the fine structure which has not been observed and determined by a detector with a certain energy resolution.
Since the Daya Bay measured spectrum is taken as the reference spectrum in our simulation, we are interested in the scale of unobserved fine structure with an 8\% energy resolution detector. Therefore, the summation spectrum we use is smeared out with 8\% energy resolution and 26 bins. Then we compare the original jagged spectrum with the smeared spectrum to estimate the scale of fine structure which is unobserved by Daya Bay.
The summation spectrum we generated is based on the cumulative yield datasets from the Evaluated Nuclear Data File (ENDF) \cite{Brown:2018jhj, ENDF} and Joint Evaluated Fission File (JEFF) \cite{JEFF}. Total absorption spectroscopy (TAS) nuclear structure data is included where available. Our spectrum using ENDF yield data is identical to reference \cite{Dwyer:2014eka} but with TAS data replacing previous structure data where available. Unless noted, all results are based on ENDF cumulative yields.

\begin{figure*}[!htbp]
\centering
\includegraphics[scale=0.5]{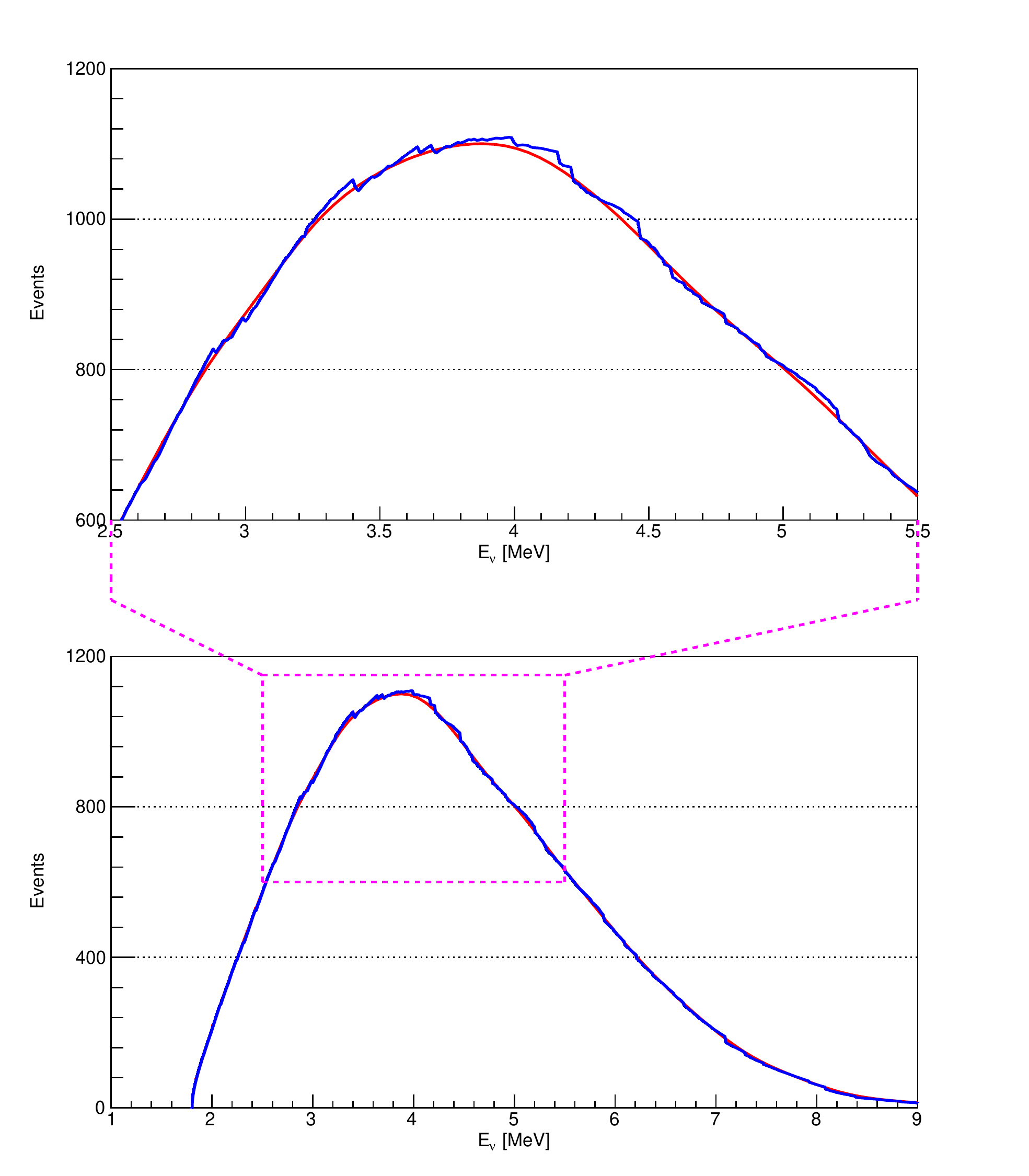}
\caption{Comparison of ``the (unoscillated) spectrum with fine structure'' to the ``smooth (unoscillated) spectrum after smearing''. The red curve corresponds to the expected smooth spectrum (the smeared out spectrum with 8\% energy resolution). The blue curve shows the original jagged spectrum without smearing.}
\label{fig:structure_compare_DB}
\end{figure*}

\begin{figure*}[!htbp]
\centering
\includegraphics[scale=0.6]{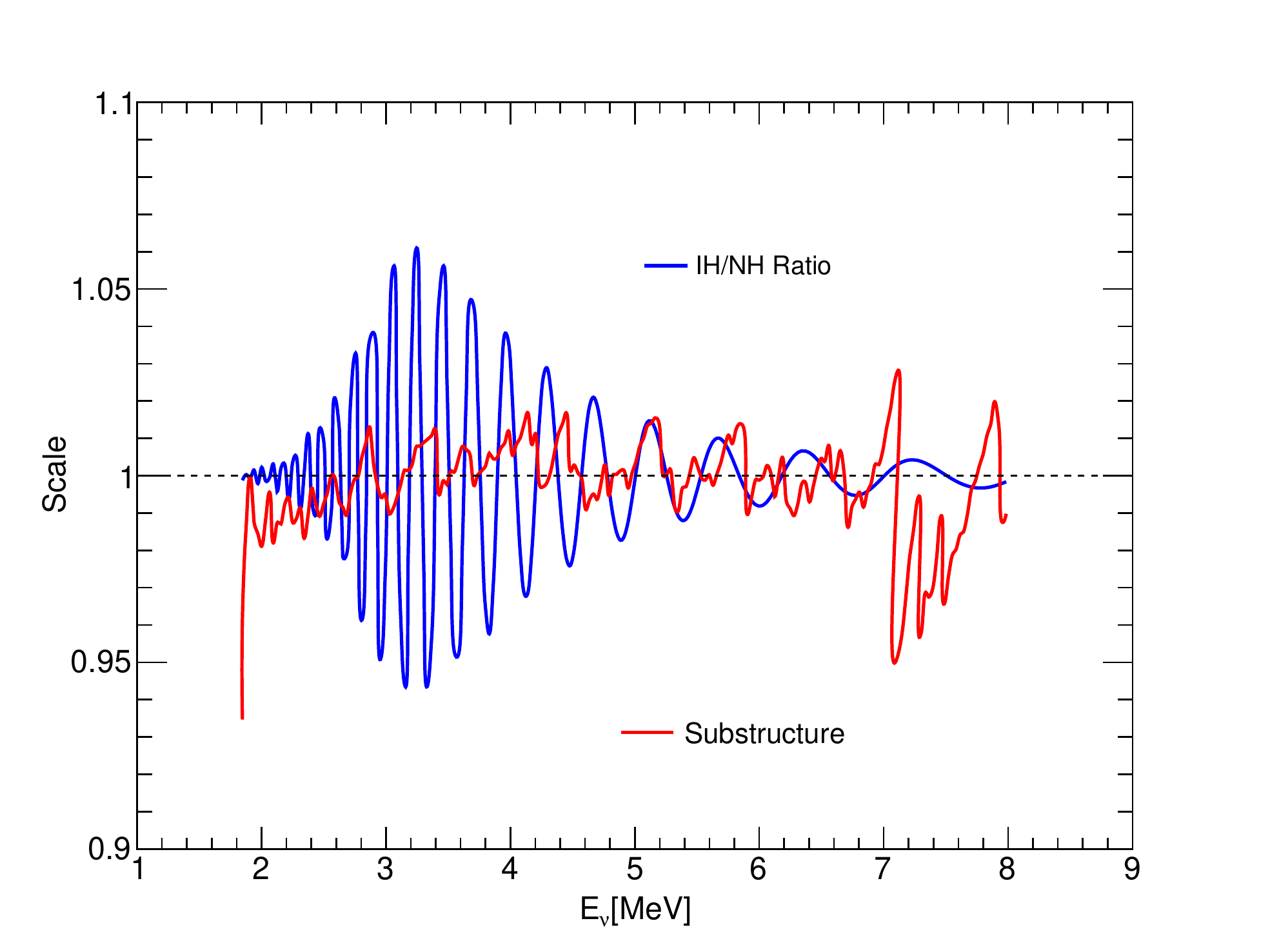}
\caption{Comparison of our estimated scale of fine structure (red) with the ratio between IH and NH (blue).}
\label{fig:structure_compare_IHoverNH}
\end{figure*}

Fig. \ref{fig:structure_compare_IHoverNH} shows that the scale of substructure could be large at high energy ($E_\nu$ $>$ 7 MeV), but is relatively small in the range of $E_\nu$ = 2 to 6 MeV, compared with the ratio of spectra corresponding to the IH and NH. Since the neutrino MH signal in MBRO experiments is mainly from this energy region, the sensitivity of such an experiment is not expected to be affected by the undetermined fine structure in the spectrum. The following subsections will investigate this more quantitatively.

\subsubsection{Fine Structure as Additional Shape Uncertainty}

Thus far, we have measured the scale of the fine structure at each point in the spectrum. However, each bin needs a single value treated as additional shape uncertainty because the same number applies across each bin. Due to the bins are sufficiently narrow, the middle value is used. We treat the deviation from 1 at the middle of each bin as an additional shape uncertainty.
This allows the true spectrum to go up or down, rather than fixing the saw-tooth structure to be same as the summation predictions. This is to account for the possibility that the true spectrum may very well be different from the spectrum obtained from the summation method as we are attempting to account for many possible saw-tooth structures.
There is a clear excess at low energy in the smooth spectrum. This is likely due to the effect of energy resolution causing events from higher energies to be detected in lower energy bins due to the sharp increase in rate from 1.8-3 MeV. Since we are treating the scale of fine structure as a shape uncertainty, this can be left in and will only make the result more conservative.
Fig. \ref{fig:structure_compare_IHoverNH} shows that the scale of
(unobserved) substructure is relatively small in the range of $E_\nu$
= 2 to 6 MeV, compared with the ratio of spectra corresponding to IH
and NH.
However, we also need to consider the shape uncertainty from our reference spectrum (a smoothed spectrum based on the Daya Bay measurement).

\begin{figure*}[!htbp]
\centering
 \includegraphics[scale=0.6]{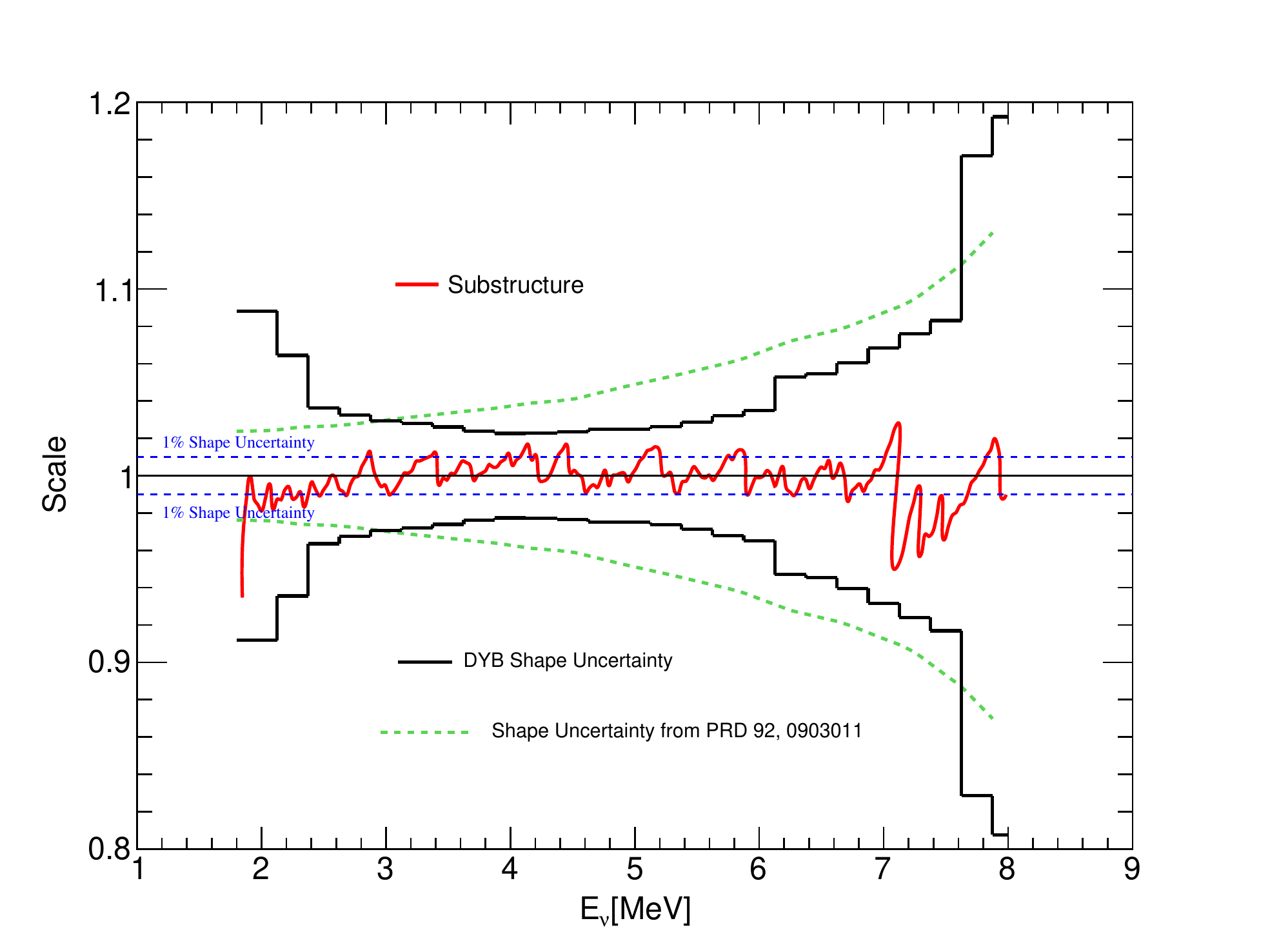}
\caption{Comparison of our estimated scale of fine structure (red) with the current shape uncertainty (black) of the measured Daya Bay spectrum \cite{An:2017osx}. We also plot the expected shape uncertainty (1\%) in the literature \cite{Yifang, An:2015jdp, Wang:2016vua} (blue dashed lines) as well as the smoothed and symmetric uncertainty from  \cite{Capozzi:2015bpa} as a comparison (the dashed green curve).}
\label{fig:structure_compare_currentDB_and_001shapeEr}
\end{figure*}

Fig. \ref{fig:structure_compare_currentDB_and_001shapeEr} shows that according to our estimation, the scale of fine structure which Daya Bay fails to measure is smaller than the current spectrum uncertainties of the Daya Bay measurement in the range $E_\nu = $ 2 to 8 MeV. The red curve representing scale of unmeasured fine structure does not exceed the shape uncertainty of the Daya Bay measurement \cite{An:2017osx}.
In our simulations, we consider both the shape uncertainties due to the uncertainties of the Daya Bay measurement, and also additional shape uncertainties due to fine structure. We sum these two different kinds of shape uncertainties in quadrature to calculate the total shape uncertainties of each bin. Specifically, in Eq. (\ref{chi2}), $\sigma_{s,i}^2 = \sigma_{\text{sub},i}^2 + \sigma_{\text{DB},i}^2$. We also calculate a result using the conventional assumption of 1\% shape uncertainty summed in quadrature with the scale of fine structure in each bin.

\subsection{The Results of Our Simulations}
\begin{figure}[!htbp]
\centering
\includegraphics[scale=0.6]{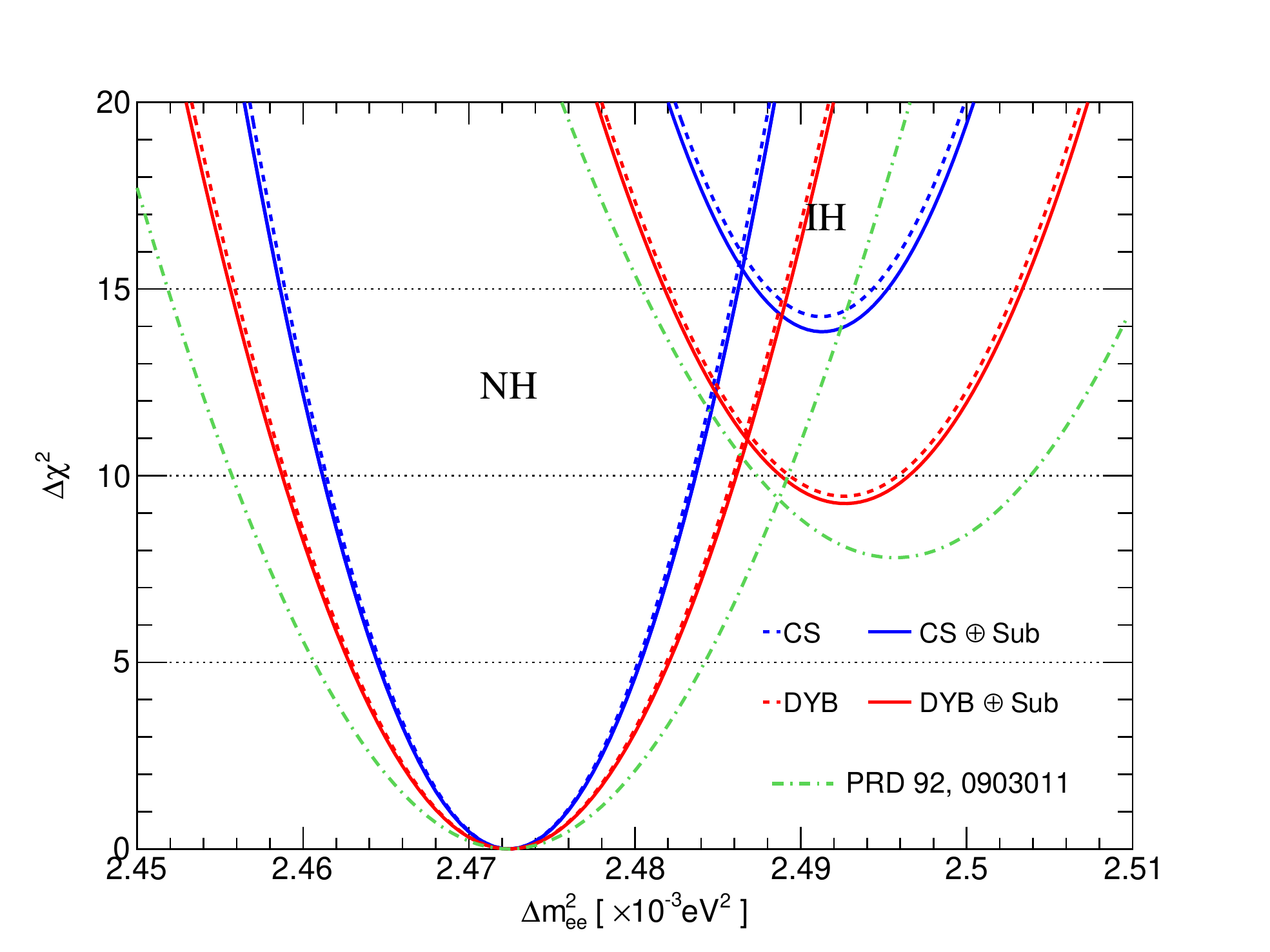}
\caption{Sensitivity of MH resolution using shape uncertainty from fine structure (solid lines) and without (dashed lines). NH is assumed to be the true MH. The curves on the left side correspond to the fitting of NH and curves on right side correspond to the assumption of IH (false MH). The blue curves represent the conventional assumption of 1\% shape uncertainties (named as ``CS'' in the legend) for each bin; The red curves correspond to intrinsic uncertainties from the Daya Bay measurement (named as ``DYB'' in the legend). The green curves are the results based on shape uncertainties reported in \cite{Capozzi:2015bpa}.}
\label{fig:compare_sensitivity_conventional}
\end{figure}

Fig. \ref{fig:compare_sensitivity_conventional} shows the comparison of our results with the estimated sensitivities of different considerations of shape uncertainties. The dashed blue curve represents the sensitivity corresponding to the conventional assumption of 1\% shape uncertainty \cite{Yifang, An:2015jdp}, while the solid blue curve shows the combination of 1\% shape uncertainty and additional shape uncertainty due to fine structure.
Because we use the Daya Bay measurement as our reference spectrum, the corresponding uncertainties of such a spectrum \cite{An:2016srz} are taken into account. The red curves in Fig. \ref{fig:compare_sensitivity_conventional} show the same result with the Daya Bay shape uncertainty replacing the assumption of 1\% uncertainty constant over energy.
The minima of the dashed and solid curves are close to each other, meaning that the consideration of fine structure does not significantly affect neutrino MH resolution. The difference between the assumption of 1\% shape uncertainty and Daya Bay spectrum shape uncertainty is much larger than the difference between the results with and without fine structure.
The result based on uncertainties reported in \cite{Capozzi:2015bpa} is also shown as a comparison and is the worst result of the three, which has not considered fine structure but just the 1 $\sigma$ uncertainty bands estimated in reference \cite{Dwyer:2014eka}\footnote{
The authors of reference \cite{Capozzi:2015bpa} do not consider the large scale structure or fine structure. They provide conservative results which are based on the estimated uncertainties of the summation calculation in reference \cite{Dwyer:2014eka}. Nevertheless, as discussed in section \ref{sec2}, the summation method cannot provide well-defined uncertainties. Moreover, we believe that the large scale structure would not affect the resolution of neutrino MH. Thus in our simulations we estimate the scale of fine structure and treat it as additional shape uncertainties.}.
Reference \cite{Capozzi:2015bpa} has also studied the potential impact of shape uncertainties. Our simulations based on Daya Bay and 1\% shape uncertainty suggests a better sensitivity than the result based on the discussion from reference \cite{Capozzi:2015bpa}, as our suggested shape uncertainties are smaller, which are based on the estimation of potential fine structure
\footnote{Please note that the sensitivity represented by the green curves in Fig. \ref{fig:compare_sensitivity_conventional} is better than the exact result in reference \cite{Capozzi:2015bpa}, since we have not considered detector nonlinearity. We want to focus on the discussion of potential structure within the reactor neutrino spectrum, thus nonlinearity is not discussed in this article.}.

With the current Daya Bay spectrum uncertainty taken into account, the sensitivity of our simulation is given by $\Delta \chi^2$ = 9.254 (9.452) with (without) shape uncertainty from fine structure. With the conventional assumption of 1\% shape uncertainty \cite{Yifang, An:2015jdp}, we find $\Delta \chi^2$ = 13.856 (14.262) with (without) fine structure.
The effect on these results from fine structure is small when compared to the difference between different assumptions of shape uncertainty. Additionally, because our treatment of fine structure is conservative, we are over-estimating its impact.

\begin{figure}[!htbp]
\centering
\includegraphics[width=0.47\textwidth]{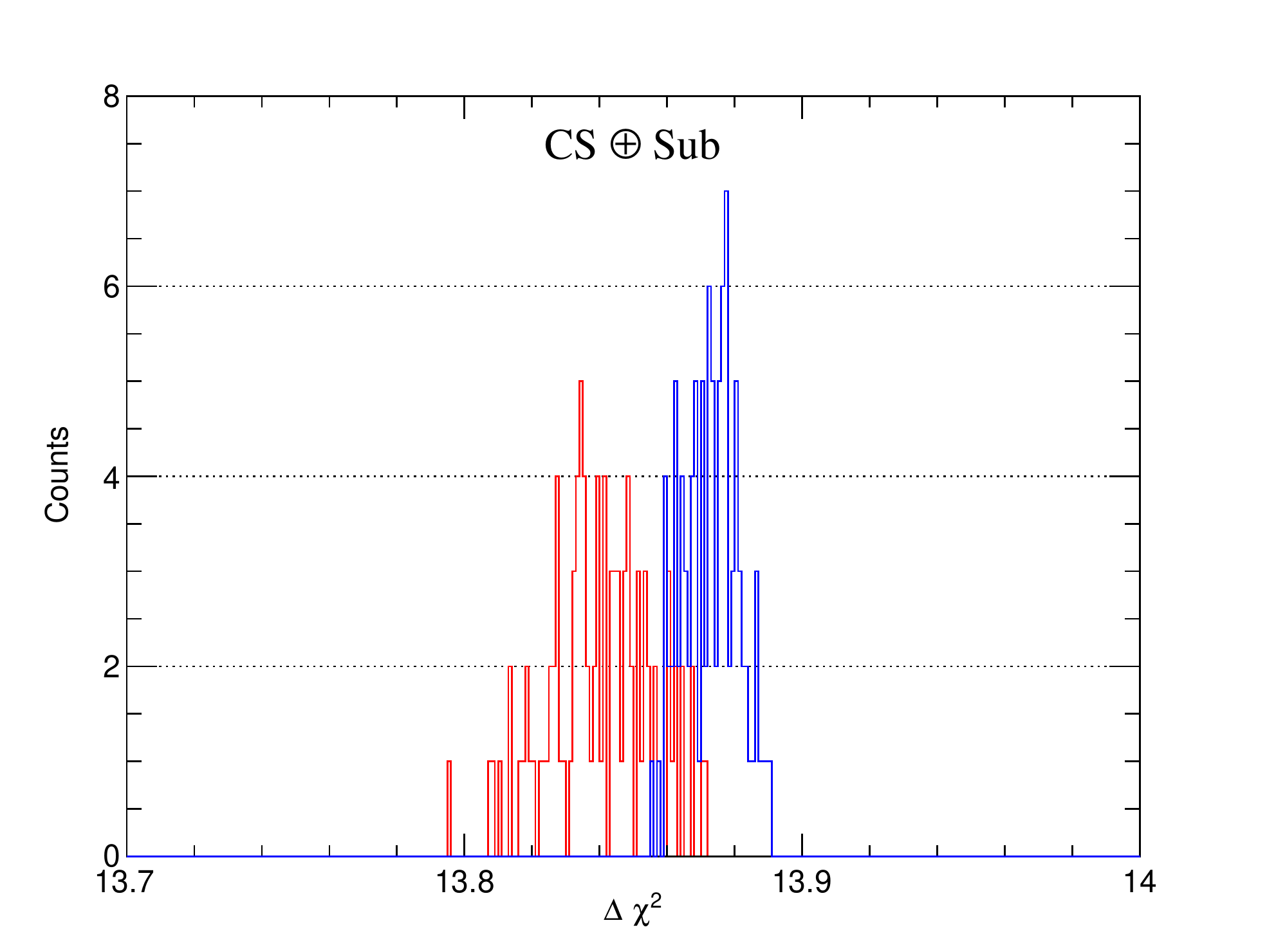}
\includegraphics[width=0.47\textwidth]{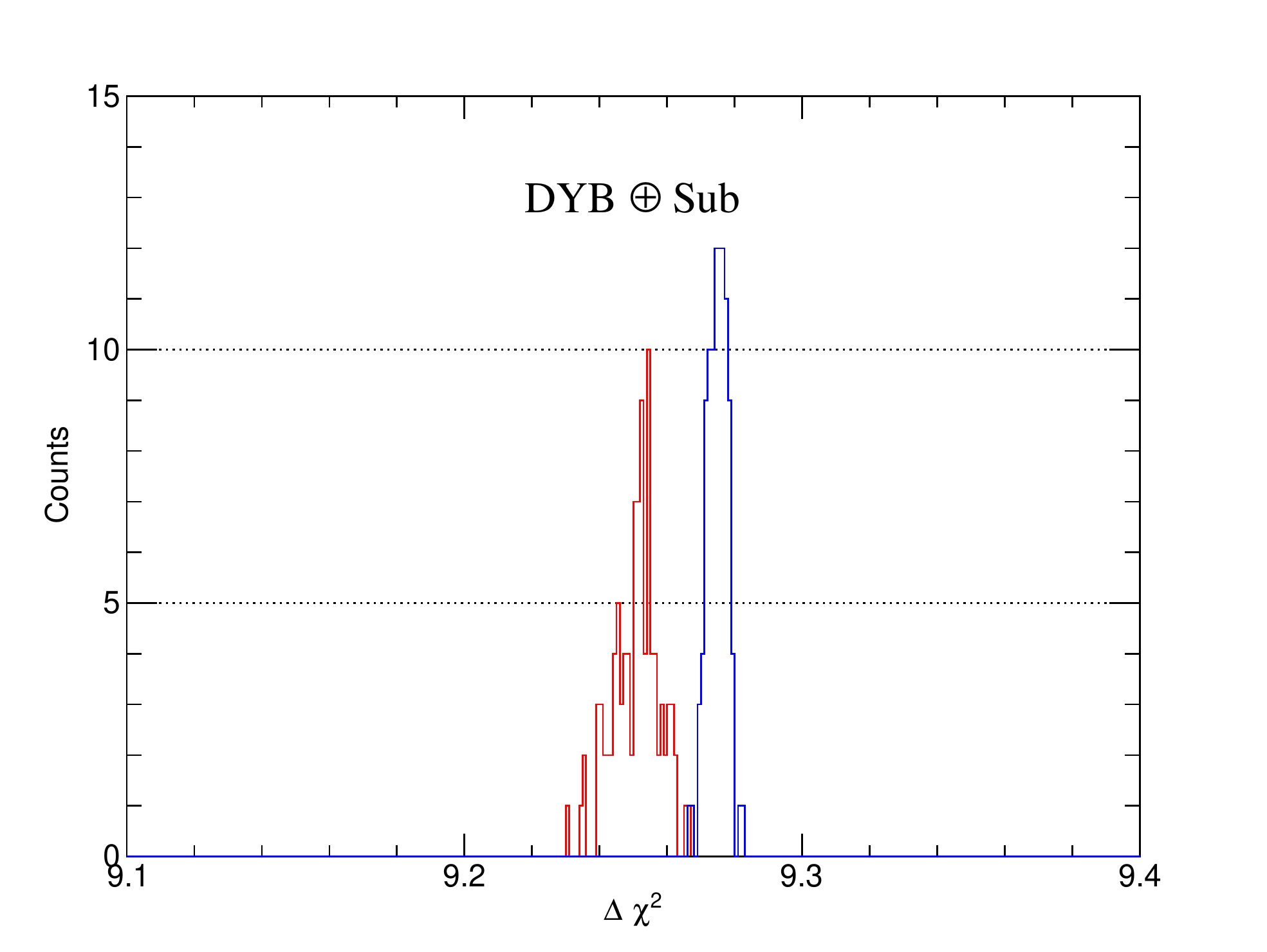}
\caption{Sensitivities from spectra with inputs varied within their uncertainties based on ENDF (red) and JEFF (blue). The left panel shows results based on 1\% shape uncertainty and the right shows results based on Daya Bay shape uncertainties.}
\label{fig:1000samples_shapeEr_sensitivities}
\end{figure}

In order to account for possible variation in the spectrum, we generate 100 spectra based on both the JEFF and ENDF databases with all inputs varied randomly within their uncertainties. Fig. \ref{fig:1000samples_shapeEr_sensitivities} shows the sensitivities of MH resolution based on the random samples generated from the ENDF (red) and JEFF (blue) datasets.
The left panel shows results based on the conventional assumption of shape uncertainty (1\%), while the right panel shows results based on Daya Bay shape uncertainties. However, both panels show that different summation spectra give similar $\Delta \chi^2$ values.
The spread in results from each alternate spectrum is far narrower than the spread between results with and without shape uncertainty from fine structure. Bias in the underlying data is not contained in the published uncertainties, but the minimal effect from varying the spectra indicates that the true spectrum is unlikely to give results which will appreciably diminish the sensitivity of MH resolution.

\subsection{Estimated Fine-Structure from Summation Spectra from the Literature}
\begin{figure*}[!htbp]
\centering
\includegraphics[scale=0.6]{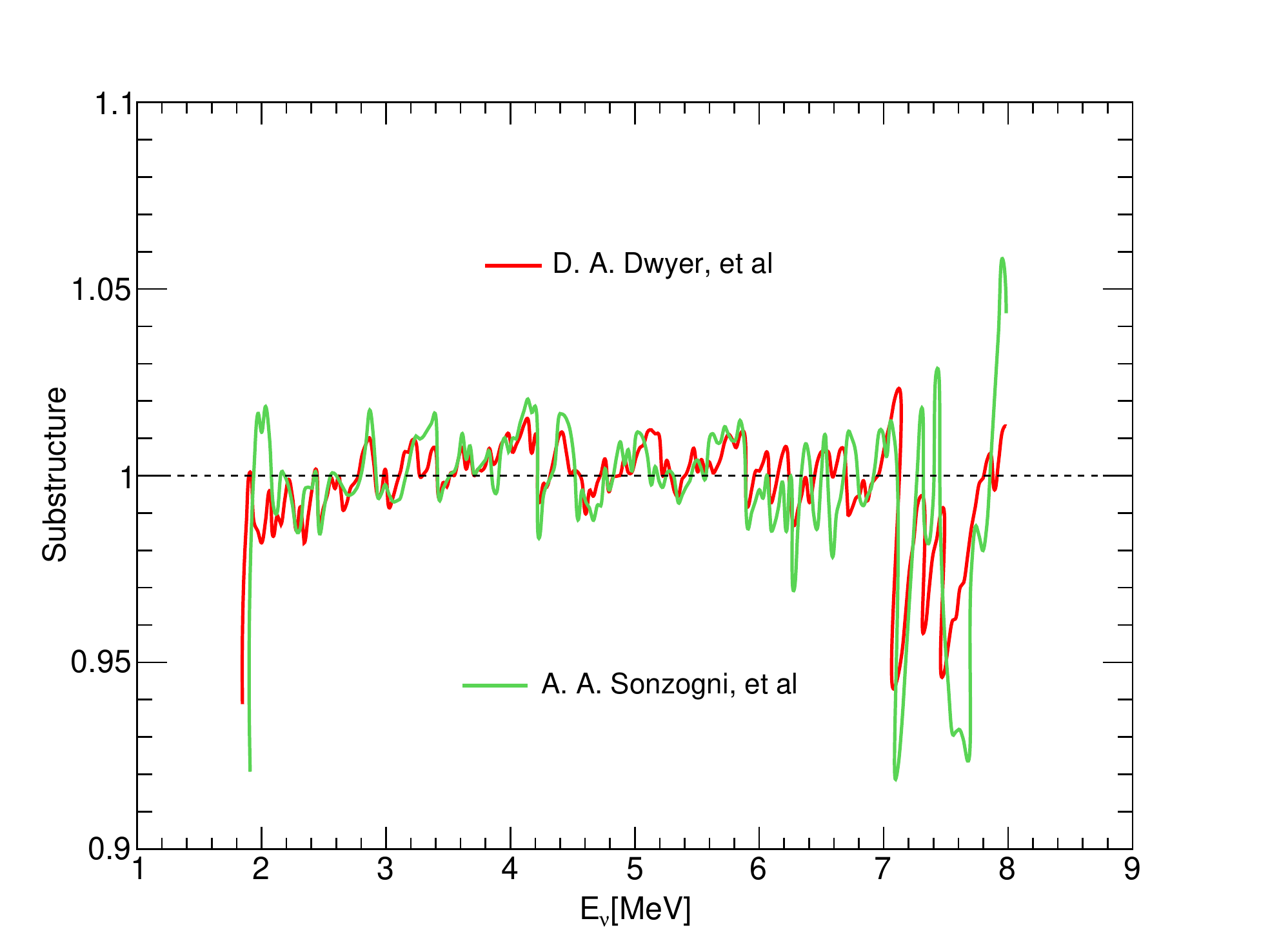}
\caption{The estimated scale of fine structure from two additional summation spectra generated by D. A. Dwyer \cite{Dwyer:2014eka} (green curve) and A. A. Sonzogni \cite{Sonzogni:2017voo} (red curve).}
\label{fig:compare_Dan_and_Sonzogni}
\end{figure*}

In this section, we repeat the analysis using two spectra from references \cite{Dwyer:2014eka, Sonzogni:2017voo} based on the summation method. As employed in the previous simulations, we use the Daya Bay measured spectrum as a reference spectrum. We then smear out the summation spectra with an energy resolution of 8\%/$\sqrt{E/MeV}$ and estimate the scale of (unobserved) fine structure based on the ratio of the original jagged spectrum to the smeared out spectrum.
Again, even the exact shape of these two summation spectra are so different, we find that the corresponding scales of (unobserved) fine structures are similar and smaller than 1 to 2\% over the low energy range ($E_\nu$ = 2 - 6MeV). The comparison of these two fine structures is shown in Fig. \ref{fig:compare_Dan_and_Sonzogni}.

\begin{figure}[!htbp]
\centering
\includegraphics[width=0.47\textwidth]{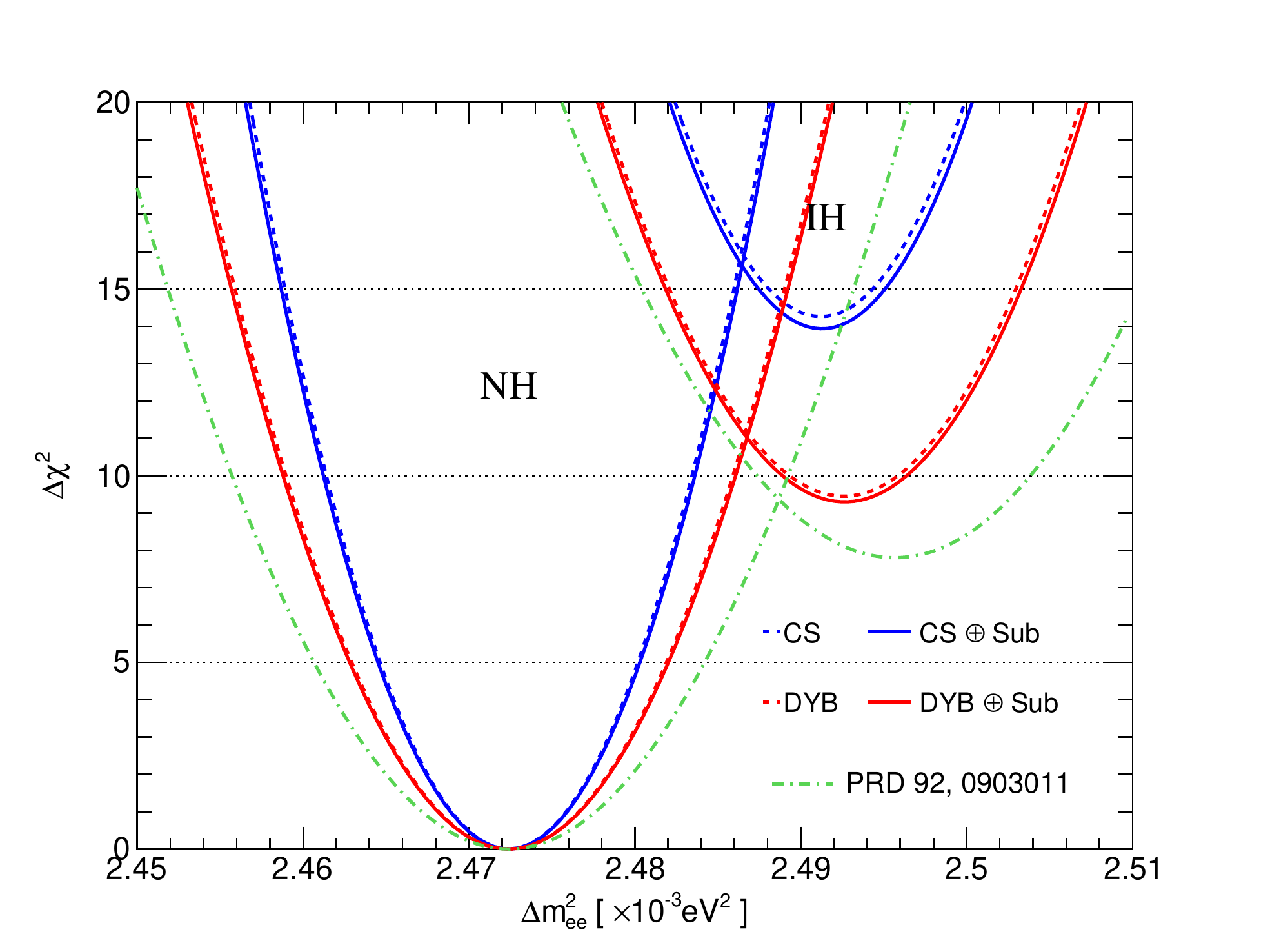}
\includegraphics[width=0.47\textwidth]{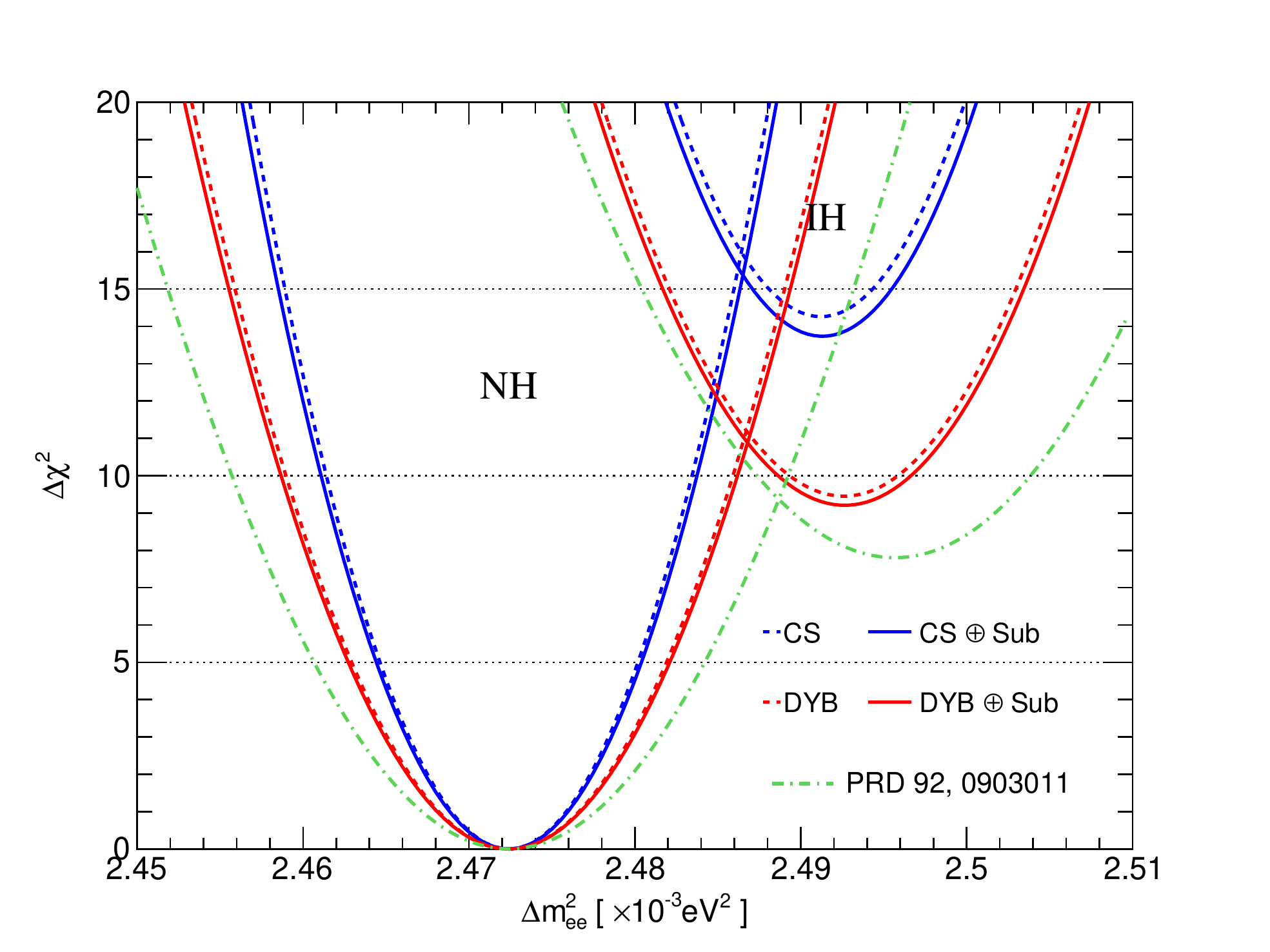}
\caption{Left: Results with fine structure based on the spectrum in reference \cite{Dwyer:2014eka};
Right: Results with fine structure based on the spectrum from reference \cite{Sonzogni:2017voo}.}
\label{fig:compare_Sonzogni_sensitivity_conventional}
\end{figure}

As we would expect from the similarity between the scale of fine structure obtained from each summation spectra, the sensitivity of MH resolution is only slightly affected by these summation spectra.
Results based on the estimated fine structures from references \cite{Dwyer:2014eka} and \cite{Sonzogni:2017voo} are shown in Fig. \ref{fig:compare_Sonzogni_sensitivity_conventional}.
The $\Delta \chi^2$ based on the spectrum from \cite{Dwyer:2014eka} is 9.298, while the $\Delta \chi^2$ based on the spectrum from \cite{Sonzogni:2017voo} is 9.208. Both results use Daya Bay shape uncertainty and have very similar $\Delta \chi^2$ values to that obtained from our own spectrum ($\Delta \chi^2$ = 9.254).
This suggests that the choice of summation spectra does not significantly affect the sensitivity of neutrino MH resolution, and reinforces the conclusion that introducing fine structure as additional shape uncertainty only slightly degrades sensitivity. As showed in Fig. \ref{fig:compare_sensitivity_conventional}, in both panels of Fig. \ref{fig:compare_Sonzogni_sensitivity_conventional}, the sensitivities of the conventional assumption of 1\% uniform shape uncertainties are much larger, regardless of whether we include shape uncertainty from fine structure.

\section{MH Resolution with a Near Detector}\label{sec4}
\subsection{Motivations of a Near Detector}
The results from previous sections suggest that fine structure is not likely to significantly affect the sensitivity of MH resolution. However, it is still worthwhile to examine the potential effect of a near detector on MH resolution in terms of its ability to measure fine structure. In fact, building an additional detector for MBRO experiments has been recently discussed in the literature \cite{Wang:2016vua, Ciuffoli:2012bp}. Particularly, the JUNO experiment \cite{An:2015jdp} is planning to build a near detector 30-35 m from a European pressure water reactor (EPR) of 4.6 GW thermal power, JUNO-TAO\cite{JUNO-TAO2}.
Such detector is designed to be with target mass around 1 ton and the energy resolution is expected to be $\sim$ 1.5$\%/\sqrt{E/\mathrm{MeV}}$. With such an excellent energy resolution, the JUNO-TAO detector or any other near detector could improve the measurement on the fine structure and provide a more precise reactor neutrino spectrum.

Taking the spectrum measured by a near detector as a reference spectrum will help cancel correlated systematic uncertainties, such as the reactor neutrino shape uncertainties and also the uncertainties due to the non-linear detector energy response correction \cite{2AD}. However, building an identical far and near detector is not feasible because any far detector capable of determining the MH is quite large with a unique geometry, and there will therefore be many uncorrelated uncertainties to deal with.
Additionally, if a near detector starts data taking after the far detector, this could introduce additional uncorrelated uncertainties. In our simulations, the proposed near detector is not used to cancel the correlated uncertainties, and we do not implement any model of detector nonlinearity.
Our analysis focuses instead on determining the benefit of a near detectors ability to measure fine structure with a constant shape uncertainty assigned to both detectors over all energy.

Similar to the proposal of JUNO-TAO, we assume a small detector with 1 ton target-mass, located 33 meters from the reactor core. In reality, there are several reactor cores located at different positions. Therefore, one small detector located near just one reactor will observe a slightly different spectrum than the original far detector due to differing fuel compositions between reactors. Such differences will generate more uncorrelated uncertainties. However, in this paper, we focus on the ideal case: one powerful reactor, one large far detector and one small near detector.

\subsection{The Energy Resolution of the Near Detector}
\begin{figure}[!htbp]
\centering
\includegraphics[scale=0.6]{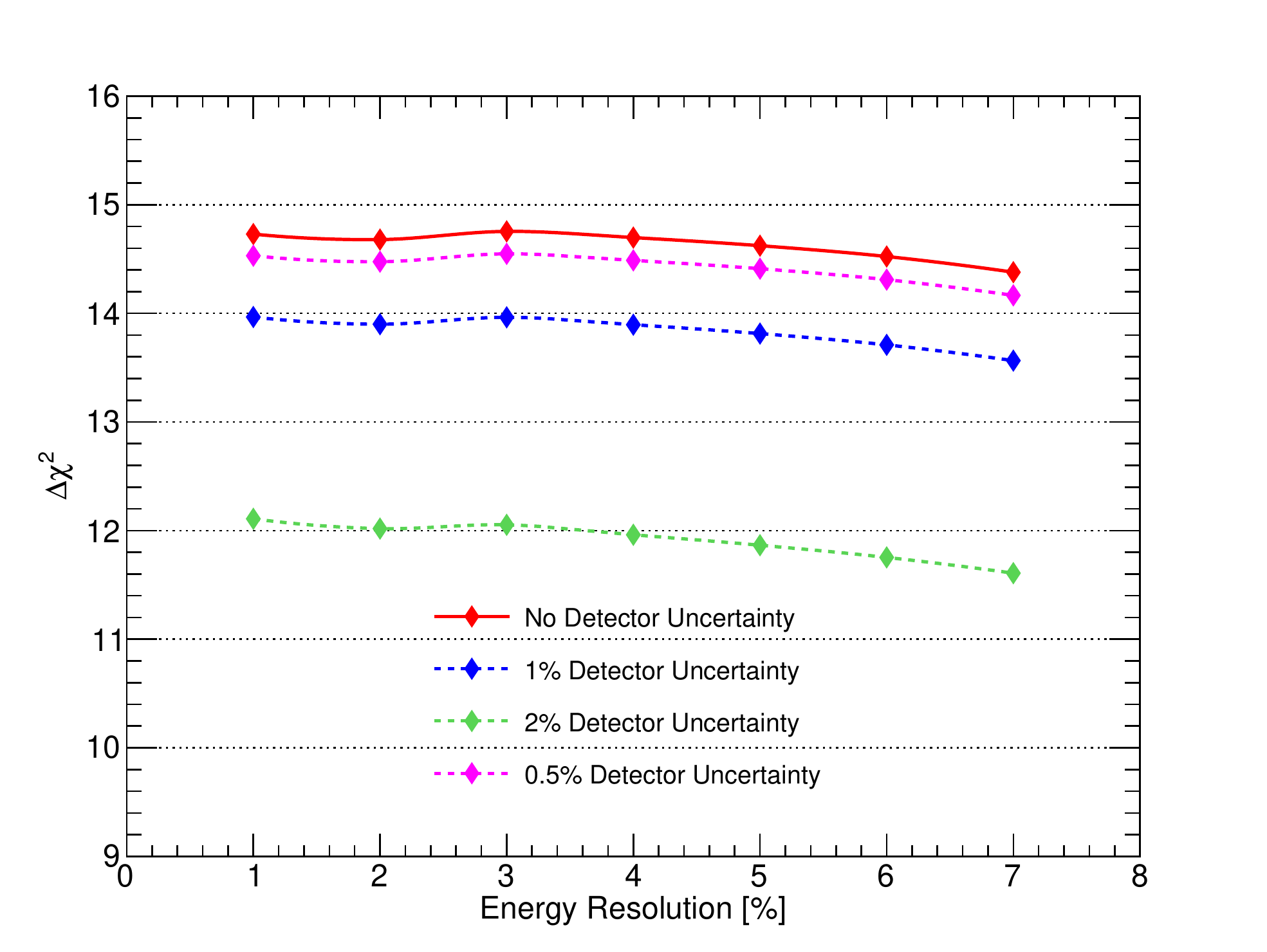}
\caption{Sensitivity of MH resolution as the energy resolution of the near detector increases and for different detector uncertainty of ND. The near detector provides the reference spectrum, which is smoothed according to the energy resolution of the ND. Similar to our previous simulations based on the Daya Bay measurement in section \ref{sec3}, the detector uncertainty of ND is treated as part of the shape uncertainty in the resolution of MH. }
\label{fig:fine_structure_ShapeEr_ND_chi2}
\end{figure}

We firstly assume the systematic uncertainties of the near detector (ND) are same as the far detector: 1\% detector-related uncertainty \cite{Yifang, An:2015jdp, Wang:2016vua}. The baseline is set to be just 33 meters. We assume sufficient exposure time for this mass and baseline that statistical uncertainty is negligible.
Without loss of generality, we assume our summation spectrum (ENDF with additional TAS data) is the true spectrum which will be measured by a near detector. We use this spectrum to estimate the unobserved fine structure in the near detector. This spectrum is used to calculate the values of $T_i$ in Eq. (\ref{chi2}), which is the expected event rates measured in the far detector with 3\% energy resolution.
The near detector (measured) spectrum is taken as the reference spectrum, which is smoothed according to several different energy resolutions and used to calculate the values of $F_i$ in Eq. (\ref{chi2}). In other words, we fit the far detector data based on the spectrum measured by the ND.
The results of this approach, shown in Fig. \ref{fig:fine_structure_ShapeEr_ND_chi2}, shows that as we would expect from results reported in the previous section detector uncertainty is more important to MH sensitivity than energy resolution. The lack of impact on sensitivity from energy resolution of ND, indicates that MH sensitivity is not appreciably affected by the fine structure predicted by current summation predictions. On the other hand, the detector uncertainty of ND could be important as it represents the uncertainty of the reference spectrum, which would be treated as shape uncertainty in our analysis of the data collected in far detector. More details about the detector uncertainty of ND will be discussed in Appendix.

\section{Conclusion}\label{sec5}
In order to resolve the neutrino MH with MBRO experiments, it was argued that a precise measurement of the fine structure of the reactor antineutrino spectrum should be achieved.
To study the impact of fine structure on the sensitivity to neutrino MH, we have used different summation spectra to estimate the potential scale of the un-observed fine structure in the Daya Bay measurement and also a proposed near detector. All of these spectra predict similar scales of fine structure, which are small over the low energy range ($E_\nu$ = 2 - 6MeV) even with 8\%/$\sqrt{E/\mathrm{MeV}}$ energy resolution. Our simulations show that the impact of such fine structure on MH resolution is insignificant, especially when compared to the uncertainties of the reference spectrum. Compared to the unobserved fine structure, the detector uncertainty of the proposed near detector (or the systematic uncertainties of the Daya Bay measured spectrum) are more important to the resolution of the neutrino MH.

\appendix
\section{The Detector Systematic Uncertainty of Near Detector}
Subsection 4.2 clearly shows that concerning the potential to resolving neutrino MH, the systematic uncertainties of ND could have larger impact than the ND energy resolution. This is because the systematic uncertainties of the ND represent the uncertainties of the reference spectrum.
The uncertainties of the detector energy nonlinearity are energy dependent. Such uncertainties of the ND could lead to energy dependent uncertainties of the reference spectrum, which are then propagated to the shape uncertainties of the analysis in FD.
\footnote{Moreover, the nonlinearity of the FD itself is also expected to have significant impacts on the neutrino MH
  resolution, as it could distort the observed spectrum and thus is crucial.}
However, in this article, the focus of our study is on the uncertainties due to the undetermined shape of reactor flux and the ways in which an extra detector provide a precise reference spectrum. The sources of the nonlinearity and methods to reduce corresponding uncertainties are not discussed. More details about the studies of nonlinearity can be referred to reference \cite{An:2015jdp}. On the other hand, the benefits of an extra detector in canceling the uncertainties of nonlinearity are discussed in reference \cite{2AD}. However, different with our assumption, the authors of that reference assume the original medium baseline detector and the proposed extra detector are identical and the correlated uncertainties can be canceled. We believe that the correlation of the nonlinearity between the ND and FD require more careful and detailed studies, which is beyond the scope of this article.

\begin{figure}[!htbp]
\centering
\includegraphics[scale=0.6]{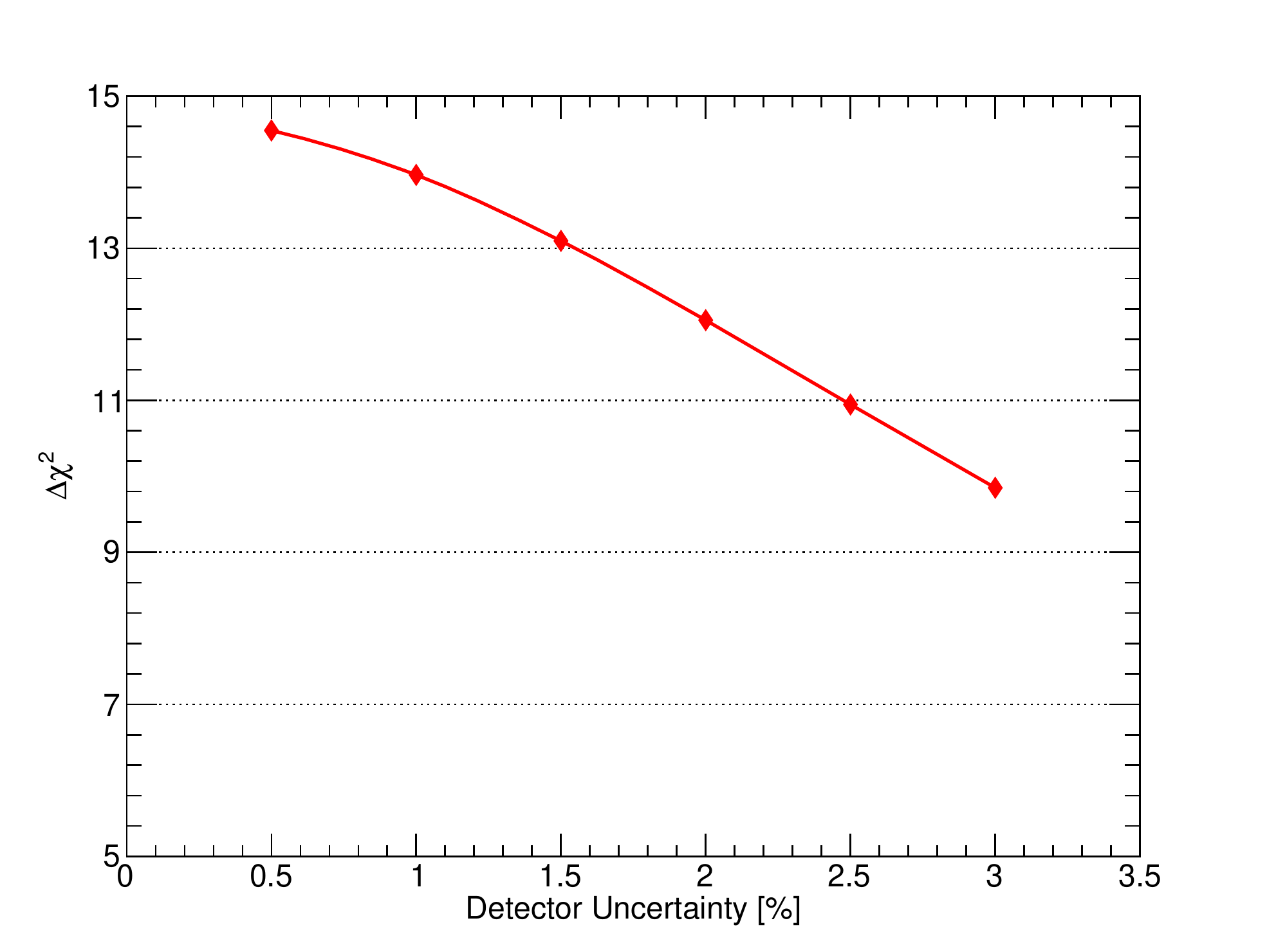}
\caption{MH Resolution sensitivity for different detector uncertainties and a constant energy resolution of 3\%.}
\label{fig:fine_structure_DetEr_ND_chi2}
\end{figure}

Here, we just assume the overall detector uncertainties are consistent with all energy bins in the ND measurement, similar to the assumption of shape uncertainties in references \cite{Yifang, An:2015jdp, Wang:2016vua}. Please keep in mind that the detector uncertainties of ND would give rise to additional shape uncertainties of the measurement in FD, since the ND uncertainties correspond to the uncertainties of reference spectrum in our simulations. Fig. \ref{fig:fine_structure_DetEr_ND_chi2} shows the sensitivity to neutrino MH vs the detector uncertainty. The latter is treated as additional shape uncertainty in our simulation of the MH resolution at far detector. Since the energy resolution of the ND barely makes impact on the MH resolution, we just assume an ND energy resolution of 3\% in Fig. \ref{fig:fine_structure_DetEr_ND_chi2}.

Fig. \ref{fig:fine_structure_DetEr_ND_chi2} shows that the sensitivity of MH resolution is strongly dependent on the systematic uncertainties of the ND measurement. Our analyses show that for MH resolution, the intrinsic uncertainties of the reference spectrum could be more important than resolving the fine structure of the reactor flux, since the scale of the unobserved fine structure is expected to be smaller than 1\%. In the future, if a near detector is built, the corresponding systematic uncertainties of nonlinearity, detection efficiency, background estimation, etc could be important.

\section*{Acknowledgement}
The authors thank D.A. Dwyer, T.J. Langford, Xin Qian, Jiajie Ling, Jarah Evslin, Yu Feng Li and Suprabh Prakash for informative discussions and suggestions.

\bibliographystyle{elsarticle-num}
\bibliography{reference_reactor.bib} 

\end{document}